\documentclass[twocolumn,pre,floatfix,superscriptaddress,showpacs]{revtex4}
\usepackage{graphicx}
\usepackage{times}
\usepackage{color}
\usepackage{amsmath}
\usepackage{epstopdf}
\def\E{{\rm e}}
\def\I{{\rm i}}
\begin{document}
\title{Dynamics of defects in active nematics}
\author{L.~M. Pismen}
\affiliation{Department of Chemical Engineering and Minerva Center for Nonlinear Physics of Complex Systems, Technion -- Israel Institute of Technology, Haifa 32000, Israel}
\begin{abstract}
Defect dynamics in a thin active nematic layer is studied by asymptotic matching of solutions in the defect core and the far field. The analysis is facilitated by the correspondence between the 2D nematic and complex scalar field models. Self-propulsion and topological interactions are identified as the primary drivers of the defect motion, surpassing the influence of both passive backflow and active flow induced by other defects.
\end{abstract}
\pacs{87.10.Ca, 83.10.Ff, 47.57.Lj, 83.80.Xz}
 \maketitle

Long-scale dynamics of ordered media is dominated by the motion of topological defects \cite{Bray}. While the nature of defects is well understood based on topology of the order parameter field \cite{Mermin,kleman}, their dynamics strongly depends on the character of dissipative processes in particular cases, and poses great difficulties to analytical studies \cite{book}. Dissipative dynamics of point defects driven exclusively by topological interactions has been studied in the context of the 2D complex scalar field, XY-model, dislocations in non-equilibrium patterns, and smectics C, all of them leading to identical mathematical formulation \cite{Pleiner,PR90,Neu,xy}, as well as in the 3D setting of nematic liquid crystals \cite{hhog}. Later numerical \cite{jul02,zumer} and analytical \cite{tim} studies elucidated the influence of backflow on the dynamics of nematic defects, which has some common features with the action of induced flow on disclinations in non-equilibrium patterns \cite{p92}. 

Recently, much attention, largely driven by biophysical applications, has been attracted to \emph{active} ordered media \cite{jkpj,review}. Very recent numerical simulations \cite{march,julia} revealed rich dynamics dominated by creation, motion, and annihilation of defects. It is the aim of this Letter to attain analytical insight into defect dynamics in non-equilibrium systems of this kind, which balance gain and dissipation. The principal tool, similar to near-equilibrium applications \cite{book}, is perturbation analysis based on matching solutions in the defect core, where the absolute value of the order parameter varies, with the far field where only phase or orientational field is relevant. The analysis is facilitated by reducing the 2D nematic model to the dynamics of a complex scalar field. As a result, the structure and dynamics of half-integer-charged nematic defects is projected on the dynamics of integer-charged vortices.   

I consider a thin flat layer with parallel anchoring on both bounding interfaces. Under these conditions, the nematic order parameter can be assumed constant across the layer, leaving only a dependence on in-plane coordinates $\mathbf{r}=(x,y)$, and 2D formulation is appropriate. 
The 2D nematic order parameter is a traceless symmetric tensor with the components $Q_{ij}=\rho (2n_in_j-\delta_{ij})$, where $\rho $ is its absolute value, $\mathbf{n}=(\cos \theta ,\, \sin \theta)$ is the unit director, $\theta$ is the orientation angle, and $\delta_{ij}$ is the Kronecker delta. An equivalent more convenient representation is
\begin{equation}
\mathbf{Q} = \left(\begin{array}{cc}p & q \\ q & -p \end{array} \right)
\equiv \rho  \,\left(\begin{array}{cc}\cos 2\theta & \sin 2\theta \\ \sin 2\theta & -\cos 2\theta \end{array} \right),
\end{equation} 
with $\rho =(p^2+q^2)^{1/2}$. The nematic energy per unit thickness is expressed as ${\cal F} = \int {\cal L} \,d^2 \mathbf{r}$ with the 2D Landau--de Gennes Lagrangian \cite{degennes}
\begin{align}
{\cal L} = & -\frac{\alpha}{4} Q_{ij}Q_{ij} +
\frac{\alpha}{16}\left(Q_{ij}Q_{ij} \right)^2 \notag \\
& + \frac{\kappa_1}{2} \left|\partial_i Q_{ij} \right|^2 + \frac{\kappa_2}{4}
\sum_{ijk}\left(\partial_iQ_{jk}\right)^2. \label{eq:LQdef}
\end{align}
The coefficients at the algebraic terms are rescaled to the common value $\alpha$ to ensure $\rho =1$ in the homogeneous nematic state; the cubic term vanishes identically in 2D. The number of distinct Frank energy terms and elastic constants $\kappa_1, \, \kappa_2$ reduces in 2D from three to two. 
The Lagrangian can be rewritten in terms of $p$ and $q$ as
\begin{align}
& {\cal L} =  -\frac{\alpha}{2} \left(q^2 + p^2\right)+
 \frac{\alpha}{4} \left(q^2 +p^2\right)^2 + \label{eq:pq} \\
&  \frac{\kappa_1}{2}\left[\left( p_x + q_y \right)^2 
+ \left( q_x - p_y \right)^2 \right]
 + \frac{\kappa_2}{2}\left[ p_x^2 +
p_y^2 + q_x^2 + q_y^2 \right]. \notag
 \end{align}

Relaxation to equilibrium follows gradient dynamics governed by the variational equation of the form $\partial_t \mathbf{Q} = -\Gamma\delta {\cal F}/\delta \mathbf{Q}$ with the mobility coefficient $\Gamma$. Scaling time $t$ by $(\alpha\Gamma)^{-1}$ and length by the healing length $\xi = \sqrt{(\kappa_1 + \kappa_2)/\alpha}$, and varying Eq.~\eqref{eq:pq} yields the dynamic equations in a particularly simple form

\begin{equation}
  p_t =   \nabla^2 p + p - (p^2+q^2)p ,  
  \quad 
 q_t = \nabla^2 q + q - (p^2+q^2)q  ,
 \label{eq:pqt} \end{equation}
where $\nabla^2$ is the 2D Laplacian. These equations reduce to a single equation for the complex variable $\chi=p+\I q =\rho \E^{\I\vartheta}$ identical to the equation of dissipative dynamics of a vortex of unit charge in a complex scalar field \cite{book}: 
\begin{equation}
 \chi_t = \nabla^2 \chi + \chi - |\chi|^2 \chi .
\label{eq:evolps}
\end{equation}
Consider a static defect with the charge $\pm \frac 12$. In its far field, i.e. at distances from the core exceeding the healing length (taken here as unity), $\rho \to 1$ and $\vartheta =2\theta=\pm \phi$, where $\phi$ is the polar angle.  
The solution in the defect core can be obtained using the \emph{ansatz} $\chi =\rho (r)\E^{\pm \I \phi}$, leading to the equation defining the dependence of the scalar order parameter $\rho =|\chi|$ on the radial coordinate $r$: 
\begin{equation}
\rho _{rr}+r^{-1}\rho _r-r^{-2}\rho + (1-\rho ^2)\rho  =0. 
\label{eq:evolQ}
\end{equation}
This defines the well-known short-scale core structure $\rho (r)$, identical to that of a superfluid vortex of unit charge. The asymptotics of this solution are $\rho \to a r$ at $r \to 0$, where the constant $a$ is computed numerically as $a \approx 0.583$, and $\rho (r) \asymp 1 - \frac 12 r^{-2}$ at $r \gg 1$. 

We will now explore both active and passive flow induced by defects. Neglecting the viscous anisotropy, the flow field $\mathbf{u}(\mathbf{r},z)$ is determined by the Stokes equation
\begin{equation}
\mathbf{u}_{zz} = - \mathbf{F}(\mathbf{r}), \quad 
\mathbf{F}= \eta^{-1}\left[\nabla P - \nabla \cdot \left(\sigma^\mathrm{(p)}+ \sigma^\mathrm{(a)}\right)\right], 
\label{eq:vdef}
\end{equation}
where $\eta$ is viscosity, $\nabla$ is the 2D gradient operator, $P$ is pressure, $\mathbf{\sigma}^\mathrm{(p)}$ is the passive elastic stress, and $\mathbf{\sigma}^\mathrm{(a)}= \zeta \alpha \mathbf{Q}$ is the active stress with the activity parameter $\zeta$. In the standard case of no-slip boundary conditions at the bounding planes $z=0, \, z= h$, there is a parabolic velocity profile $\mathbf{u}(z)=\frac 12\mathbf{F} z(h-z)$ and the velocity averaged across the layer is $\mathbf{U}(\mathbf{r})=\frac{1}{12}h^2 \mathbf{F}$. The averaged velocity of an incompressible fluid in a layer of constant thickness $h$ can be expressed through the stream function as $\mathbf{U} = \nabla \times \Psi$. Taking the curl of averaged Eq.~\eqref{eq:vdef} to eliminate pressure yields the equation of $\Psi(\mathbf{r})$. We write it in the dimensionless form with the length scaled by $\xi$, time by $(\Gamma\alpha)^{-1}$, and stress by $\alpha$:
\begin{equation}
\nabla^2 \Psi = \gamma \Phi , \qquad 
\Phi = \nabla \times \left[\nabla \cdot \left(\sigma^\mathrm{(p)}+ \sigma^\mathrm{(a)}\right)\right], 
\label{eq:stokes}
\end{equation}
where $\gamma =\frac{1}{12}(h/\xi)^2 (\Gamma\eta)^{-1}$ is the dimensionless hydrodynamic mobility.

We compute first the contribution of the active stress. Taking $\sigma=\sigma^\mathrm{(a)}=\zeta \mathbf{Q}$, we express the inhomogeneity in Eq.~\eqref{eq:stokes} for defects with the charge $\pm \frac 12$ as   
\begin{align}
\Phi^\mathrm{(a)}_+ &= - \zeta \sin \phi\left(\rho _{rr}+r^{-1}\rho _r-r^{-2}\rho \right) = \zeta \rho (1-\rho ^2) \sin \phi ,\notag
 \\
\Phi^\mathrm{(a)}_- &= - \zeta \sin 3\phi\left(\rho _{rr}-3r^{-1}\rho _r+3r^{-2}\rho \right) ,
\label{eq:PhiQm}
\end{align}
where the last expression for $\Phi^\mathrm{(a)}_+$ is written using Eq.~\eqref{eq:evolQ}. In a $+ \frac 12$ defect, activity generates within the defect core a force oriented along the ``comet tail", leading to a normally oriented dipolar vorticity source. A $- \frac 12$ defect has a different structure with 3-fold symmetry, and a sextuplet vorticity source is generated instead. The lubrication approximation remains applicable in the core region, provided the healing length is much larger than the layer thickness, $\xi \gg h$, which implies, at comparable hydrodynamic and rotational viscosities, $\gamma \ll 1$. The stream function for a $+ \frac 12$ defect can be presented as $\Psi^\mathrm{(a)}_+= - \gamma \zeta \psi(r)\sin \phi$ where $\psi(r)$ satisfies
\begin{equation}
\psi _{rr}+r^{-1}\psi _r-r^{-2}\psi + (1-\rho ^2)\rho  =0. 
\label{eq:evolv}
\end{equation}
The obvious solution is $\psi(r) \equiv \rho(r)$. In the far field of a defect where $\rho$ approaches unity, the stream function reduces to 
$\Psi^\mathrm{(a)}_+(r,\phi) = -\gamma \zeta\sin \phi\, (1 - \frac 12 r^{-2})$. This yields a dipole flow field decaying as $r^{-3}$ at $r \to \infty$.
The total active flow field is a superposition of flow induced by all extant defects. 

The passive elastic stress, 
$\sigma^\mathrm{(p)}_{ij}= - \partial_j Q_{kl}\partial \mathcal{L}/\partial(\partial_i Q_{kl})$, is expressed in the adopted units as
\begin{equation}
\sigma^\mathrm{(p)}_{ij}= -2\left(\partial_i p\partial_jp  + 
 \partial_i q\partial_j q +\kappa_1\xi^{-2}\delta_{ij}\epsilon_{kl}
 \partial_k p\partial_l q \right),
\label{eq:passig} \end{equation}
where $\delta_{ij}$ is the Kronecker delta and $\epsilon_{kl}$ is the 2D antisymmetric matrix.  Only the first term contributes to the inhomogeneity in Eq.~\eqref{eq:stokes}, which is expressed as $\Phi^\mathrm{(p)}(r,\phi) =\mp 2 \sin 2 \phi f_\mathrm{p}(r)$ with
\begin{equation}
f_\mathrm{p}(r)=\rho _{rr}^2+\rho _{r}\rho _{rrr}-r^{-2}\rho_r^2 -r^{-3}\rho\rho_r +r^{-4}\rho^2 .
\label{eq:PhiQp}
\end{equation}
This reduces to $\Phi^\mathrm{(p)}(r) =\mp 2 r^{-4}\sin 2 \phi$ in the far field, while $\Phi^\mathrm{(p)}(0)=0$. The far field stream function solving Eq.~\eqref{eq:stokes} is
\begin{equation}
 \Psi^\mathrm{(p)}(r,\phi) = \pm 2\gamma r^{-2}\ln (r/ C)\sin 2\phi, 
\label{eq:intrp}
\end{equation}
where the indefinite constant $C$ has to be evaluated by matching with the defect core. This defines a quadrupole flow that decays with the distance  as $r^{-3}\ln r$, i.e. logarithmically slower than active flow. For both active and passive flow, the power-law decay persist in the far field, in spite of the strong momentum transfer to the confining walls. This is a common feature of hydrodynamic interactions in suspensions mediated by the pressure field, which is a consequence of mass conservation in incompressible fluids \cite{diamond}.

The defect velocity under combined action of orientation gradients and flow induced by other defects, as well as self-induced active flow, is computed by asymptotic perturbation analysis assuming that the defect core structure is only weakly perturbed. To account for Galilean invariance, the time derivative should be replaced by the corotational substantial derivative 
\begin{equation}
{\cal D}_{ijkl}Q_{kl} =(\partial_t + \mathbf{U} \cdot \nabla)Q_{ij} + \frac{\Omega}{2} (\epsilon_{ki} Q_{kj}- \epsilon_{ik} Q_{kj} ),
\end{equation}
where $\Omega= -\nabla^2\Psi= -\gamma\Phi$ is the vorticity pseudoscalar.

It is advantageous to transform Eq.~\eqref{eq:evolps}, complemented by the advective and rotational terms, to the comoving corotating frame:
\begin{align}
 (\mathbf{v}-\mathbf{U}_s) \cdot \nabla p &-\omega\,\mathbf{r} \times\nabla p  + \nabla^2 p 
 \notag \\  
 & + p \left(1 - p^2-q^2 \right) +\gamma\Phi q=0,  
  \label{eq:ptof} \\  
(\mathbf{v}-\mathbf{U}_s) \cdot \nabla q & -\omega\,\mathbf{r} \times\nabla q + \nabla^2 q
 \notag \\  
 & +  q(1 - p^2-q^2) -\gamma\Phi p =0 ,
 \label{eq:qtof} \end{align}
where $\mathbf{v}, \omega$ are, respectfully, the translational and rotational velocities of the defect, so far unknown. Note that rotation is a non-trivial effect, since the orientation field around the defect lacks circular symmetry. These equations can be combined as 
\begin{equation}
(\mathbf{v}-\mathbf{U}_s) \cdot \nabla \chi -\omega\, \mathbf{r} \times\nabla \chi + \nabla^2 \chi + \chi(1 -\I\gamma\Phi - |\chi|^2 )  =0.
\label{corev}\end{equation}
The flow velocity induced by other defects, which is constant across the core, does not perturb the core structure, and is eliminated by the transformation to the comoving frame. The remaining variable advective term contains the self-induced velocity $\mathbf{U}_s$.  The instantaneous velocity of a defect in a long-scale quasistationary approximation is the vector sum of the translation velocity \textbf{v} induced at its current location by orientation gradient due to other defects and the  flow velocity due to other defects. The latter, however, is much weaker than the former, as it much faster decays with the distance: both the active and passive flow decay as $r^{-3}$, while the orientation gradient in the far field is proportional to $r^{-1}$. 

Further on, we rescale $\mathbf{v} \to \varepsilon \mathbf{v}$, $\gamma \to \varepsilon \gamma$, $\omega \to \varepsilon \omega$ and expand $\chi$ in the book-keeping small parameter $\varepsilon$: $\chi=\chi_0+\varepsilon\chi_1+\ldots$. The zero-order function is $\chi_0=\rho(r)\E^{\pm\I \phi}$, where $\rho(r)$ verifies Eq.~\eqref{eq:evolQ}. The first-order equation can be written in a compact form
    \begin{equation}
   {\cal H}(\chi_1,\overline \chi_1)+ \mathcal{I}( \mathbf{r})=0,
    \label{gl1m}     \end{equation}
containing the inhomogeneity 
   \begin{equation}
  \mathcal{I}( \mathbf{r}) =( \mathbf{v}-\mathbf{U}_s)\cdot \nabla \chi_0 -\omega\, \mathbf{r} \times\nabla \chi_0  - \I\gamma\Phi\chi_0
     \label{gl1inh}     \end{equation}
and the linear operator  
   \begin{equation}
   {\cal H}(\chi_1,\overline \chi_1) = 
        \nabla^2 \chi_1 + (1 -2|\chi_0|^2)\chi_1 -\chi_0^2 \overline \chi_1,
     \label{L1}     \end{equation}
where the overline denotes the complex conjugate. This operator is self-conjugate, and has three eigenfunctions $\varphi(r,\phi)$ with zero eigenvalue: the two vector components of $\nabla \chi_0= \E^{\pm\I \phi}\mathbf{W}(r)$, where
  \begin{equation}
 \mathbf{W}(r) =\rho_r \left\{ \begin{array}{c} \cos \phi 
 \\  \sin \phi \end{array} \right\} \pm 
 \frac{\I\rho}{r} \left\{ \begin{array}{c} \sin \phi \\ -\cos \phi \end{array}\right\},
\label{eigfun} \end{equation}
that correspond to the translational degrees of freedom in the plane, and the eigenfunction $\I \chi_0$ corresponding to the rotational degree of freedom.  

The translation velocity is determined by the solvability condition of Eq.~\eqref{gl1m}, which requires the inhomogeneity to be orthogonal to the eigenfunctions with zero eigenvalue. The solvability condition is computed \cite{book} in a circle of radius $r_0$ large compared to the core size but small on the far field scale, i.e. $1 \ll r_0 \ll \varepsilon^{-1}$: 
\begin{align}
& \mbox{Re}\left\{ \int_0^{r_0}r\,dr \int_0^{2\pi} \overline \varphi \,\mathcal{I}(r,\phi)\, d\phi \right. \notag \\ & \left. + 
r_0 \int_0^{2\pi} ( \overline \varphi \partial_r \chi_1 
    - \chi_1 \partial_r \overline \varphi)_{r=r_0}\, d\phi \right\} =0.
\label{solcond0}  \end{align}

Since $\nabla \chi_0$ has polar symmetry, only inhomogeneities with the same symmetry contribute to the area integral in Eq.~\eqref{solcond0}. The only component of $\mathbf{U}_s$ yielding a non-vanishing contribution is the active self-induced flow term in a $+ \frac 12$ defect: 
\begin{align}
&  \int_0^{r_0} r\,dr \int_0^{2\pi}  \mathbf{\overline{W}}(r)\left(
 (\nabla \times \Psi^\mathrm{(a)}_+) \cdot  \mathbf{W}(r) \right) d\phi 
  \notag \\ 
& =  \pi \gamma\zeta 
\left\{ \begin{array}{c} -1 \\ \I \end{array}\right\} 
\int_0^{\infty} \rho\rho_r \left(\frac {\rho}{r} +\rho_r\right)dr.
\label{solcond1} \end{align}
Since the last integral converges, the upper limit has been extended to infinity; its value $a_1 \approx 0.36$ is computed using the numerical solution of Eq.~\eqref{eq:evolQ}. Since the $y$-component of this vector expression is purely imaginary, the non-vanishing contribution comes only from the $x$-component, i.e. is directed along the "comet tail" of the defect.

The contribution of the rotational term $-\I \gamma \Phi \chi_0$ also comes only from the self-induced active flow in a $+ \frac 12$ defect, and here again only the $x$-component is real:
\begin{align}
 - \gamma \zeta\int_0^{r_0} r\,dr & \int_0^{2\pi}\mathbf{\mathrm{Re}(\I\overline{W}}) \Phi^\mathrm{(a)}_+ \rho\, d\phi  \notag \\ 
= &   -\pi \gamma\zeta \int_0^{\infty}  \rho^2 (1-\rho^2)dr .  
\label{solcond2}
  \end{align}
The radial integral is evaluated as $a_2 \approx 0.87$. Due to the 3-fold symmetry of the active flow in a $- \frac 12$ defect, its contribution to the solvability condition always vanishes.

The inhomogeneities generated by flow do not project on the rotational eigenfunction $\I \chi_0$, and therefore the induced rotation frequency $\omega$ should be set to zero. It remains to compute the contribution of the translational term $\mathbf{v}\cdot \nabla \chi_0$. Unlike the two above integrals, it diverges at $r \to \infty$, and therefore the respective contour integral cannot be discarded but has to be matched to the far field solution that determines a weakly distorted far field of a steadily moving defect \cite{book}:
\begin{align}
& \mathrm{Re} \int_0^{r_0} r\,dr \int_0^{2\pi}\mathbf{\overline{W}}(r)\, 
\left(\mathbf{v}\cdot \mathbf{W}(r) \right) d\phi \notag \\ 
= &  \pi \mathbf{v} 
\int_0^{r_0} \left( \frac{\rho^2}{r} + r\rho_r^2 \right)dr 
= \pi \mathbf{v} \ln \frac {r_0}{a_0}
  \label{solcondv} \end{align}
with $a_0 \approx 1.126$.

The contour integral depends on the asymptotics of the first-order solution $\chi_1$. Assuming $r_0=O(\varepsilon^{-1/2})$, we have $\rho_0(r_0)= 1 - O(\varepsilon)$. It follows that the contour integral can be expressed, to the leading order $O(\varepsilon)$, through the phase field $\vartheta$ alone.

An analytical solution can be obtained for the orientational far field of a defect propagating with a constant speed $\mathbf{v}$. Since the self-induced flow rapidly decays outside the defect core, it is sufficient to find a stationary solution of the equation of the phase $\vartheta$ in the comoving coordinate frame with the origin at the defect location:
\begin{equation}
\mathbf{v} \cdot \nabla \vartheta +  \nabla^2 \vartheta =0,
\label{cofar} \end{equation}
subject to the circulation condition $\oint \vartheta \, ds=\pm 2\pi$ along any contour surrounding the origin. Equation (\ref{cofar}) is solved \cite{prost} by introducing a univalued function dual to $\vartheta$, leading to a scale-invariant expression for the phase gradient:  
 \begin{equation}
\nabla \vartheta= \pm \frac{1}{2}\,
\E^{-(\mathbf{v} \cdot \mathbf{r})/2}\,
 \mathcal{R}\left[\mathbf{v}\, K_0 \left(\frac{vr}{2}\right) -\frac{v\mathbf{r}}{r}\,K_1\left(\frac{vr}{2}\right) \right],
  \label{ffs} \end{equation}
where $v=|\mathbf{v}|$, $K_i$ are modified Bessel functions, and $\mathcal{R}$ denotes clockwise rotation by the right angle. An additional solution satisfying Eq.~(\ref{cofar}) in a trivial way is $\vartheta_\mathrm{ext} = k\, \mathbf{v} \times \mathbf{r}$ defining an arbitrary phase gradient directed normally to the propagation direction.
Taking the inner limit of Eq.~(\ref{ffs}) at $r\to 0$, one can reconstitute the phase
\begin{equation}
\vartheta = \pm\left[ \phi  \frac{\mathbf{v} \times \mathbf{r}}{2} \, \ln  \left( \frac{vr}{4} 
       e^{\gamma_\mathrm{E}-1 }\right)\right] + k\, \mathbf{v} \times \mathbf{r},
\label{asymout}     \end{equation}
where $\gamma_\mathrm{E} \approx 0.577$ is the Euler constant. This expression can be used to determine the function $\chi_1$ entering the contour integral:
\begin{equation}
\chi_1 = \E^{\I\vartheta} - \E^{\pm\I\phi} = \mp \I \E^{\pm\I\phi} \frac{\mathbf{v} \times \mathbf{r}}{2} \, \ln  \left( \frac{vr}{4}
       e^{\gamma_\mathrm{E}-1 \pm 2k }\right) .
\label{asymch1}     \end{equation}
Using this in Eq.~(\ref{solcond0}), the contour integral is computed as
\begin{align}
 r_0 \int_0^{2\pi} (\nabla \overline \chi_0 \partial_r \chi_1 
    - \nabla \chi_1 \partial_r \nabla \overline \chi_0)_{r=r_0}\, d\phi& 
\notag \\ =-\pi \mathbf{v}\ln  \left( \frac{vr_0}{4} e^{\gamma_\mathrm{E}-1/2 \pm 2k }\right)  .&
\label{cont}     \end{align}
When the area and contour integrals are used in Eq.~(\ref{solcond0}), the auxiliary radius $r_0$ falls out, and the solvability condition takes the form 
\begin{equation}
 \mathbf{v}\left[\ln  \left( \frac{v}{4a_0} \E^{\gamma_\mathrm{E}-1/2}\right)\pm 2k \right] - 
 a_s\gamma\zeta  \widehat\mathbf{x} =0,  
\label{a0a1}  \end{equation}
where, for a $+\frac 12$ defect, $a_s=a_1 + a_2 \approx 1.23$  and $\widehat\mathbf{x}$ is the unit vector along the $x$ axis in the direction corresponding to the sign of the defect, i.e. towards its ``comet tail"; for a $-\frac 12$ defect, $a_s=0$.

This relation defines the coefficient $k$, so far indefinite. The resulting dependence on a weak phase (orientation) gradient $\nabla\vartheta_\mathrm{ext}$ for a defect with the charge $\pm 1/2$ can be written as 
\begin{equation}
 \nabla \vartheta_\mathrm{ext} = \pm \frac 12\mathcal{R} \mathbf{v} 
 \left[\ln \frac{v_0}{v} + a_s\gamma\zeta 
  \frac {\widehat\mathbf{x}\cdot \mathbf{v}}{v^2} \right],
\label{eq:motv} \end{equation}
where $v_0 = 4/a_0 \,\E^{1/2-\gamma_\mathrm{E}}\approx 3.29$. Note that, since $v=O(\varepsilon) \ll 1$, the logarithm is positive. In the absence of activity, this expression accounts for mutual attraction of oppositely charged defects and repulsion of defects of the same charge; it ceases to be exact as motion becomes non-stationary with changing defect separation but non-stationary corrections are weak \cite{PRS}. The essential role of motion is regularisation of the far field divergence of a static defect \cite{book}. An alternative phenomenological regularisation, based on the distance cut-off \cite{Pleiner} yields a logarithmic factor including the defect separation rather than velocity.

The \emph{self-propulsion} effect due to the active flow is made clear when Eq.~\eqref{eq:motv} is rewritten, neglecting its left-hand side, as 
 $ \mathbf{v}\ln (v_0/v) = - a_s\gamma\zeta$. 
Thus, in the absence of external forces, the $+\frac 12$ defect moves as if pushed by its ``comet tail" when the activity is tensile ($\zeta>0$) and in the opposite direction when the activity is contractile ($\zeta<0$). Being confined in a thin layer, the defect is effectively a ``crawler" rather than a ``swimmer", and self-propulsion is facilitated by the transfer of momentum to the confining planes. There is a striking asymmetry between positively and negatively charge defects, only the former being driven by self-induced active flow. Passive backflow does not affect the motion in the leading approximation, and rotation of the defect may be caused by higher-order corrections only. 

\begin{figure}[t]
\centering
\includegraphics[width=.45\textwidth]{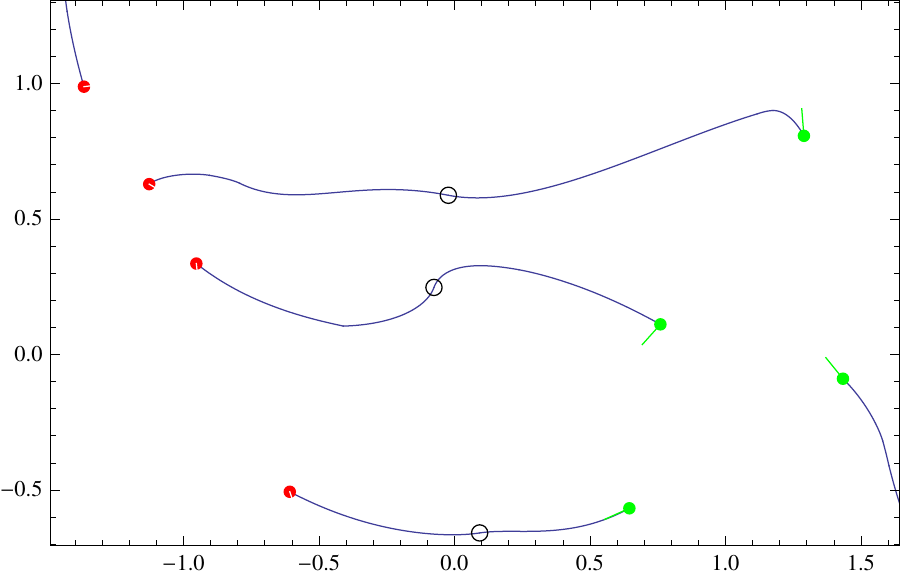}
\caption{\label{f2} (Color online) Trajectories of defects computed using Eq.~\eqref{eq:motv}, starting from 4 positive and 4 negative defects scattered randomly within the unit squares separated by the unit distance. The ``comet tails" of $+\frac 12$ defects, marked near their initial positions, are directed at random angles distributed within the intervals $(-\pi/2,\pi/2)$. The annihilation sites (the earliest below and the latest above) are marked by circles}
\end{figure}

The above analysis identifies self-propulsion and topological interactions as the primary drivers of the defect motion, surpassing the influence of both passive backflow and active flow induced by other defects. The simple formula \eqref{eq:motv} enables fast computation of defect trajectories and accumulation of statistics, without restricting to bounded or periodic domains. A simple example is shown in Fig.~\ref{f2}. Since self-propulsion remains constant while topological attraction decreases with distance, some defects escape, which happens with a higher probability as activity increases. 

Annihilation of defect pairs is compensated by their creation driven by instability of a uniformly aligned state in an active nematic. This instability, first detected in a quasi-1D confined geometry \cite{Joanny}, can be readily seen by linearising Eq.~\eqref{eq:qtof} in the vicinity of the state $p=1, \, q=0$ aligned along the $x$ axis, and substituting there $\Phi=\zeta(q_{xx}-q_{yy})$, as follows from Eq.~\eqref{eq:stokes}. This leads to the linear equation $q_t = \nabla^2q -\gamma\zeta(q_{xx}-q_{yy})$, which is absolutely unstable on all wavelengths at $\gamma|\zeta|>1$.
Creation of defect pairs is outside the scope of the present theory, but it suggests that a necessary  perturbation should have a sufficient spatial extent $L$ allowing the self-propulsion to counteract topological attraction when $L^{-1} \le O(\gamma\zeta)$. 

The above results can be compared in a qualitative way to recent simulations  \cite{march,julia}. In these simulations, a strong perturbation of the nematic field gives rise to the formation of a defect pair with the ``comet tail" of the $+\frac 12$ defect directed towards its negative counterpart. At $\zeta>0$, the resulting propulsion may turn out to be sufficiently strong to overcome the topological attraction, so that the pair separates. When applied to the dynamics of a pair of defects, Eq.~ \eqref{eq:motv}, obtained here by rigorous perturbation expansion assuming strong friction and large defect separation, differs from the simple phenomenological formula in Ref.~\cite{march} only by a logarithmic factor. The simulations were based, however, on solving the 2D Navier--Stokes equation, which, if applied to a thin layer, presumes perfect slip at the confining planes, while here the lubrication approximation is used to describe a no-slip Hele--Show geometry with strong wall friction. In the free-slip 2D setting, the momentum transfer from an active particle is impeded, leading to far-field divergences. The apparent qualitative similarity of the results, leaving the topological interactions and active flow as prevailing factors, is likely to be explained by inertial screening of far field divergences in the numerical study.

I thank Julia Yeomans for stimulating discussions, and acknowledge the support by the Human Frontier Science Program (Grant RGP0052/2009-C) and hospitality of the Isaac Newton Institute for Mathematical Sciences.

\end{document}